\documentclass[preprint]{aastex}
\shortauthors{CHUNG}
\shorttitle{RESONANT CAUSTIC EVENTS}
\newcommand{\thetae}{\theta_{\rm E}}
\newcommand{\re}{r_{\rm E}}

\newcommand{\zbar}{\bar {z}}
\newcommand{\zpbar}{\bar{z}_p}
\newcommand{\delxic}{\Delta \xi_c}
\newcommand{\delxip}{\Delta \xi_p}

\begin{document}
\title{Characterization of the Resonant Caustic Perturbation}

\author{Sun-Ju Chung}
\affil{
Korea Astronomy and Space Science Institute, Hwaam-Dong,
Yuseong-Gu, Daejeon 305-348, Korea; sjchung@kasi.re.kr
}


\begin{abstract}
Four of nine exoplanets found by microlensing were detected by the resonant caustic, which represents the merging of the planetary and central caustics at the position when the projected separation of a host star and a bounded planet is $s \sim 1$.
One of the resonant caustic lensing events, OGLE-2005-BLG-169, was a caustic-crossing high-magnification event with $A_{max} \sim$ 800 and the source star was much smaller than the caustic, nevertheless the perturbation was not obviously apparent on the light curve of the event.
In this paper, we investigate the perturbation pattern of the resonant caustic to understand why the perturbations induced by the caustic do not leave strong traces on the light curves of high-magnification events despite a small source/caustic size ratio.
From this study, we find that the regions with small-magnification-excess around the center of the resonant caustic are rather widely formed, and the event passing the small-excess region produces a high-magnification event with a weak perturbation that is small relative to the amplification caused by the star and thus does not noticeably appear on the light curve of the event.
We also find that the positive excess of the inside edge of the resonant caustic and the negative excess inside the caustic become stronger and wider as $q$ increases, and thus the resonant caustic-crossing high-magnification events with the weak perturbation occur in the range of $q \leqslant 10^{-4}$.
We determine the probability of the occurrence of events with the small excess $|\epsilon| \leqslant 3 \%$ in high-magnification events induced by a resonant caustic.
As a result, we find that for the Earth-mass planets with a separation of $\sim 2.5\ \rm{AU}$, the resonant caustic high-magnification events with the weak perturbation can occur with a significant frequency.

\end{abstract}


\section{INTRODUCTION}

Until now over 370 extrasolar planets have been discovered by various methods including the radial velocity method (Mayor \& Queloz 1995; Marcy \& Butler 1996), transit method \citep{struve52}, direct imaging \citep{stahl95}, pulsar timing analysis \citep{wolszczan92}, astrometry \citep{pravdo09}, and microlensing \citep{mao91}, and nine of them have been detected by the microlensing technique.
The microlensing signal of a planet is a short-duration perturbation on the smooth standard light curve of the primary-induced lensing event that occurred on a background source star.
The perturbation for Jupiter-mass planets lasts for a few days, while for Earth-mass planets it lasts for a few hours.
High cadence observations are thus required for the detection of extrasolar planets using microlensing.
Survey and follow-up observations for microlensing are now being carried out toward the Galactic bulge field.
The survey observations (OGLE: Udalski 2003; MOA: Bond et al. 2002) monitor a large area of sky and alert ongoing events by analyzing data in real time, while the follow-up observations ($\mu$Fun: Dong et al. 2006; PLANET: Albrow et al. 2001) intensively monitor the alerted events.

In planetary lensing, the perturbation is induced by the central and planetary caustics, which are typically separated from each other.
The strength of the perturbation depends upon the sizes of the caustics together with the finite-source effect.
The perturbation becomes strong as the sizes of the caustics increase and the finite-source effect decreases.
When the central and planetary caustics merge at the position of the projected star-planet separation normalized to the Einstein radius of the lens system, $s \sim 1.0$, the caustic size is maximized and thus the probability for the detection of a planet is the highest at $s = 1.0$.
This merging caustic is so called the $``$resonant caustic".
In fact, four of nine extrasolar planets found by microlensing (Bond et al. 2004; Gould et al. 2006; Gaudi et al. 2008; Janczak et al. 2009) have been detected by the resonant caustic.
The resonant caustic is generally much bigger than the central caustic at a given mass ratio, and thus high-magnification events caused by the resonant caustic are much less affected by the finite-source effect than those caused by the central caustic.
Thus, we can easily expect that the perturbations of the resonant caustic-crossing high-magnification events will noticeably appear on their light curves.
However, OGLE-2005-BLG-169 \citep{gould06} was a resonant caustic-crossing high-magnification event with a small source/caustic size ratio, yet the perturbation on the light curve of the event was weak and thus it was difficult to recognize the perturbation without the residual from the single lensing magnification.
In this paper, we investigate the perturbation pattern of the resonant caustic to find out why this is so.

The paper is organized as follows.
In Section 2, we briefly describe the properties of the planetary lensing.
In Section 3, we investigate the perturbation pattern of the resonant caustic and determine the probability for events with the small excess $|\epsilon| \leqslant 3\%$ of the resonant caustic high-magnification events.
We conclude the results in Section 4.

\section{PLANETARY LENSING}

Planetary lensing is described by the special case of binary microlensing with a very low mass ratio.
Because of the very small mass ratio, the planetary lensing behavior is well described by that of a single lens of the host star for most of the event duration.
In this case, the lens equation (Bozza, V. 1999; An J. H. 2005) is expressed as
\begin{equation}
\zeta = z - {1\over{\zbar}} - {q\over{\zbar -\zpbar}},
\end{equation}
where $\zeta = \xi + i\eta$ and $z = x + iy$ represent the complex notations of the source and image positions, respectively, $\zbar$ denotes the complex conjugate of $z$, $z_p$ is the position of the planet, and $q$ is the planet/star mass ratio.
Here the position of the star is at the center and all lengths are normalized to the Einstein ring radius $\thetae$ of the total mass of the lens system, i.e.,
\begin{equation}
\thetae = \sqrt {{4GM \over c^2}{\left ({1 \over D_{\rm L}} - {1 \over D_{\rm S}}\right )}}\ ,
\end{equation}
where $D_{\rm L}$ and $D_{\rm S}$ are the distances to the lens and the source from the observer, respectively.

One of two sets of disconnected caustics in the planetary lensing, $``$central caustic", is always formed close to the host star and thus the perturbation induced by the central caustic occurs near the peak of the light curve.
Events with the source star trajectory passing near the central caustic produce high-magnification events and the perturbation induced by the central caustic has generally a property of the $s \leftrightarrow 1/s$ degeneracy (Griest \& Safizadeh 1998; Dominik 1999).
The size of the central caustic as measured by the width of the cusps on the star-planet axis \citep{chung05} is expressed as
\begin{equation}
\delxic \sim {4q \over {(s - 1/s^2)}}\ .\\
\end{equation}

The other, $``$planetary caustic", is formed away from the host star and thus the perturbation induced by the planetary caustic can occur at any part of the light curve of the event.
The shape and number of the caustics depend on whether the planet lies inside ($s < 1$) or outside ($s > 1$) the Einstein ring.
Although the planetary caustic could be one or two of them depending on whether $s < 1$ or $s > 1$, the position of the planetary caustic is always located at $s - 1/s$.
The sizes of the planetary caustic for $s < 1$ and $s > 1$ \citep{han06} are respectively expressed as
\begin{displaymath}
\delxip \sim \left\lbrace \begin{array}{ll}
{3\sqrt{3q}}s^{3}/4 & \textrm{for $s < 1$} \\\\
{4\sqrt{q}s^{-2}}\left( 1 + 1/2s^{2}\right) & \textrm{for $s > 1$} \ .\\
\end{array} \right.
\end{displaymath}
The planetary caustic is always bigger than the central caustic and its size decreases more slowly than that of the central caustic as $q$ decreases, because $\delxic \propto q$ and $\delxip \propto \sqrt{q}$.

As $s \rightarrow 1.0$, the planetary caustic approaches the central caustic and finally merges with the central caustic.
The merging starts at the position close to $s = 1.0$ and the position moves away from $s = 1.0$ as $q$ increases.
During the merging, the planetary caustic part of the resonant caustic is still located at $s - 1/s$.

\section{RESONANT CAUSTIC PERTURBATION}

To investigate the perturbation pattern of the resonant caustic, we construct magnification excess maps of the planetary systems with the resonant caustic.
The magnification excess is defined by
\begin{equation}
\epsilon = {A - A_{0} \over {A_0}}\ ,
\end{equation}
where $A$ and $A_0$ are the lensing magnifications with and without a planet, respectively.

Figure 1 shows the magnification excess maps of the planetary lensing systems of various mass ratios and separations close to 1.0.
The dotted circle has a radius of $u_{0}=0.01$, which produces a high-magnification event.
In each map, the regions with blue and red-tone colors represent the areas where the excess is negative and positive, respectively.
The color in Figure 1 changes into darker scales when the excess is $|\epsilon| = 1\%,\ 3\%,\ 5\%,\ 10\%,\ \rm {and},\ 20\%$, respectively.
Considering that the masses and distances of the lens of the observed high-magnification planetary lensing events are mostly distributed in the range of $0.46\ M_{\odot} \leqslant M_{\rm L} \leqslant 0.56\ M_{\odot}$ and $1.0\ \rm {kpc} \leqslant D_{\rm L} \leqslant 3.3\ \rm{kpc}$, respectively, we assume that $M_{\rm L}=0.5M_{\odot}$, $D_{\rm L} = 2 \rm{kpc}$, $D_{\rm S} = 8 \rm{kpc}$, and the source is a main-sequence star with a radius of $R_\star = 1.0\ R_\odot$.
Then, the corresponding angular Einstein radius is $\thetae = 1.235\ \rm{mas}$, and thus the source radius normalized to the Einstein radius is $\rho_{\star} = \theta_\star/\thetae = (R_\star/D_{\rm S})/\thetae = 4.7\times10^{-4}$.

From the map, we find that after the central caustic merges with the planetary caustic, the following features commonly appear in the perturbation pattern of the resonant caustic.
\begin{enumerate}
\item
A small-magnification-excess region inside the resonant caustic become considerably wider as $s \rightarrow 1.0$, where it occupies the region toward the planet from the caustic center and the region is maximized at $1.0 < s < 1.01$.
This is because after the merging, the fold caustics with strong positive excess inside the caustic become further away from the star-planet axis by approaching the planetary caustic part of the resonant caustic as $s \rightarrow 1.0$.
\item
A gap between the negative and positive excesses outside the resonant caustic becomes wider as $s \rightarrow 1.0$ and thus the small-excess region outside the caustic increases.
This is because after the merging, the negative excesses outside the caustic become weakened by approaching the cusps of the planetary caustic part of the resonant caustic with positive excess, as $s \rightarrow 1.0$, while the positive excesses become displaced from the caustic center by the elongation of the cusps with positive excess located at the opposite side to the planet as $s \rightarrow 1.0$.
\end{enumerate}

The light curves and residuals from the single lensing one resulting from the source trajectories presented in Figure 1 are shown in Figure 2.
In the upper part of each panel, black and red curves represent the light curves of the planetary and single lensing events, respectively.
The lower part shows the residual from the single lensing magnification.
As shown in the upper part of each panel, the caustic-crossing events passing the small-excess region around the caustic center produce high-magnification events with weak perturbations.
The perturbations are small relative to the amplifications caused by the star and thus do not appear strongly on the light curves.
Thus, the light curves of those events seemingly appear to be those of the single lensing events.
However, the resonant caustic-crossing events for $q = 5.0\times10^{-4}$ induce strong perturbations that appear noticeably on the light curves.
This is because the positive excess of the inside edge of the resonant caustic and the negative excess inside the caustic become stronger and wider as $q$ increases.
From this, we find that the resonant caustic-crossing high-magnification events with weak perturbation occur in the range of $q \leqslant 10^{-4}$.

To investigate the frequency of planetary microlensing events with weak perturbations in high-magnification events induced by the resonant caustic, we determine the probability of occurring events with the small excess $|\epsilon| \leqslant 3 \%$.
The perturbations with $|\epsilon| \leqslant 3 \%$ are not obviously apparent on the light curves of high-magnification events.
For this, we set the threshold lens-source impact parameter, $u_{0,\rm{th}}=0.01$.
The probability is determined from many source trajectories within the threshold impact parameter and the result is presented in Figure 3.
As shown in Figure 3, the probability well supports prior results, where the probability is maximized at $1.0 < s < 1.01$ and most of the probabilities occur in the range of $q \leqslant 10^{-4}$.
Unlike shown in the top panel of Figure 1, the probability for $q = 10^{-5}$ with the projected separation of $1.02 \leqslant s \leqslant 1.03$ is rather small.
This is because although the area with $|\epsilon| \leqslant 3 \%$ in the dashed circle is wide, the magnification excesses close to the star-planet axis are mostly $|\epsilon| > 3\%$ and thus possible source trajectories with $|\epsilon| \leqslant 3 \%$ are limited to the cases parallel with the star-planet axis.
In Table 1, we also present the average probability of occurring the resonant caustic high-magnification events with weak perturbation in the range of $1.0 \leqslant s \leqslant 1.02$.
Considering of the physical parameters of the lens being assumed in Section 3, the physical Einstein radius is $\re = 2.5\ \rm{AU}$ and thus the planets presented in Figure 3 and Table 1 are located around $2.5 \rm{AU}$ from the host star.
We find that for $3M_{\rm E}$ planets with a separation of $\sim 2.5 \rm{AU}$, the resonant caustic high-magnification events with weak perturbation can occur with a frequency $\sim 21 \%$.

A high-magnification event with a weak perturbation can be resolved by the pattern of the residuals from the single lensing magnification around the peak.
Fortunately, since current microlensing observations are focusing on high-magnification events and the photometric error reaches $\sim 1 \%$ at the peak of a high-magnification event, the high-magnification events with $|\epsilon| \geqslant 3 \%$ can be easily resolved.
However, the high-magnification event with $|\epsilon| < 1 \%$ shows up in the range of $q = 10^{-5}$, and in this case, it is very difficult to resolve the event using the residual pattern, even if an intensive observation is
 carried out around the peak.

\section{CONCLUSION}

We have investigated the perturbation pattern of the resonant caustic.
From this, we found that small-magnification-excess regions around the resonant caustic center are rather widely formed, and the event passing the small-excess region produces a high-magnification event with the weak perturbation despite the small source/caustic size.
This perturbation is small relative to the amplification caused by the star and thus does not show up well on the light curve.
We also found that the resonant caustic-crossing high-magnification events with the weak perturbation occur in the range of $q \leqslant 10^{-4}$, because the positive excess of the inside edge of the resonant caustic and the negative excess inside the caustic become stronger and wider as $q$ increases.
We have determined the probability of the occurrence of events with the small excess $|\epsilon| \leqslant 3 \%$ in the resonant caustic high-magnification events.
From the determination, we found that for the Earth-mass planets with a separation of $\sim 2.5\ \rm{AU}$, the resonant caustic high-magnification events with the weak perturbation can take place with a significant frequency.

\begin{figure}[t]
\epsscale{1.0}
\plotone{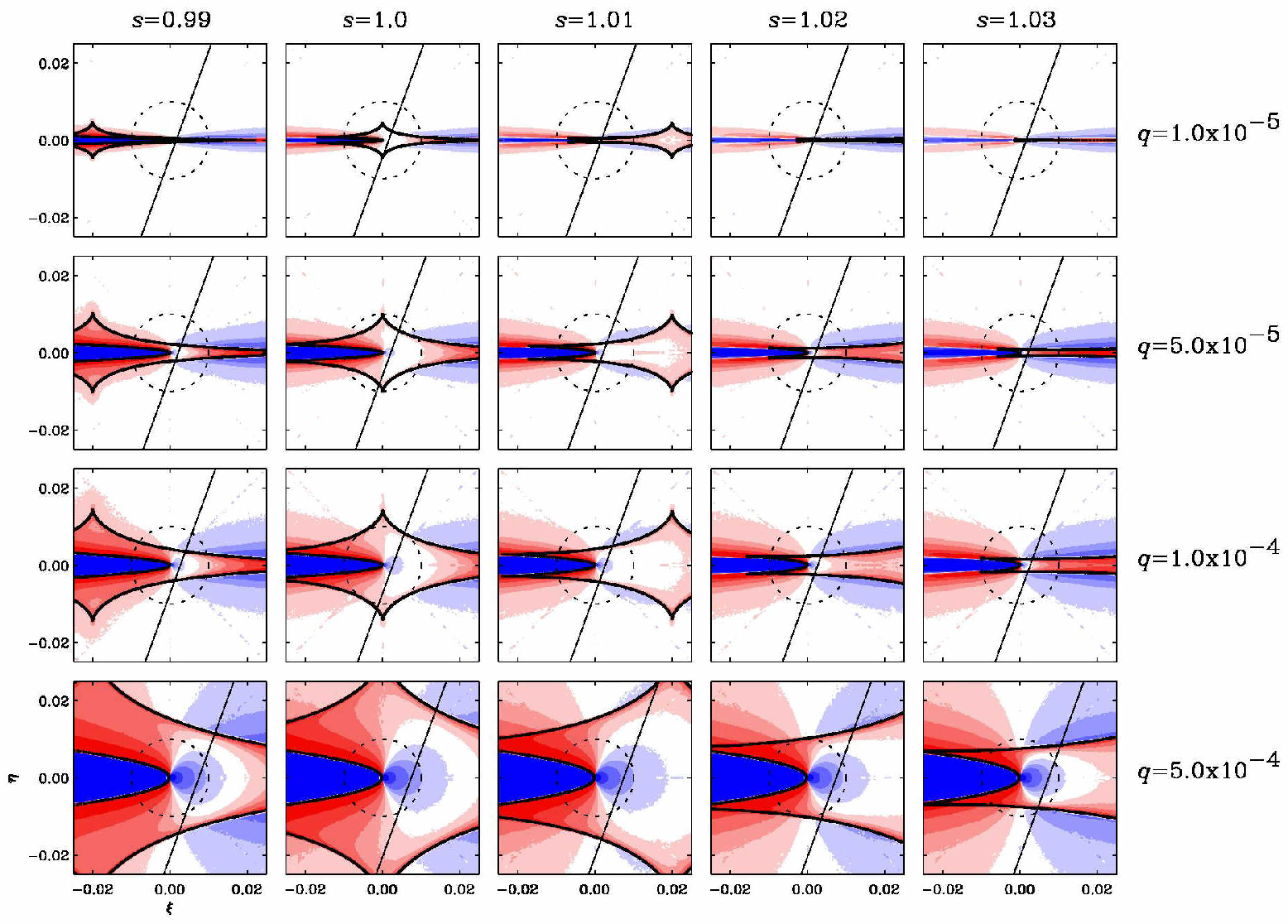}
\caption{\label{fig:one}
Magnification excess maps of the planetary lensing systems of various mass ratios and separations close to 1.0. The coordinates ($\xi, \eta$) represent the axes that are parallel with and normal to the star-planet axis and are centered at the caustic center.
In each map, the planet is located on the right and all lengths are normalized to the Einstein angular radius, $\thetae$.
The dotted circle has a radius of $u_{0}=0.01$, within which a source would generate a high-magnification event.
The regions with blue and red-tone colors represent the areas where the excess is negative and positive, respectively.
The color changes into darker scales when the excess is $|\epsilon| = 1\%,\ 3\%,\ 5\%,\ 10\%,\ \rm {and},\ 20\%$, respectively.
The straight lines with arrows represent the source trajectories, and the light curves of the resulting events are presented in the corresponding panels of Fig. 2.
}\end{figure}

\begin{figure}[t]
\epsscale{1.0}
\plotone{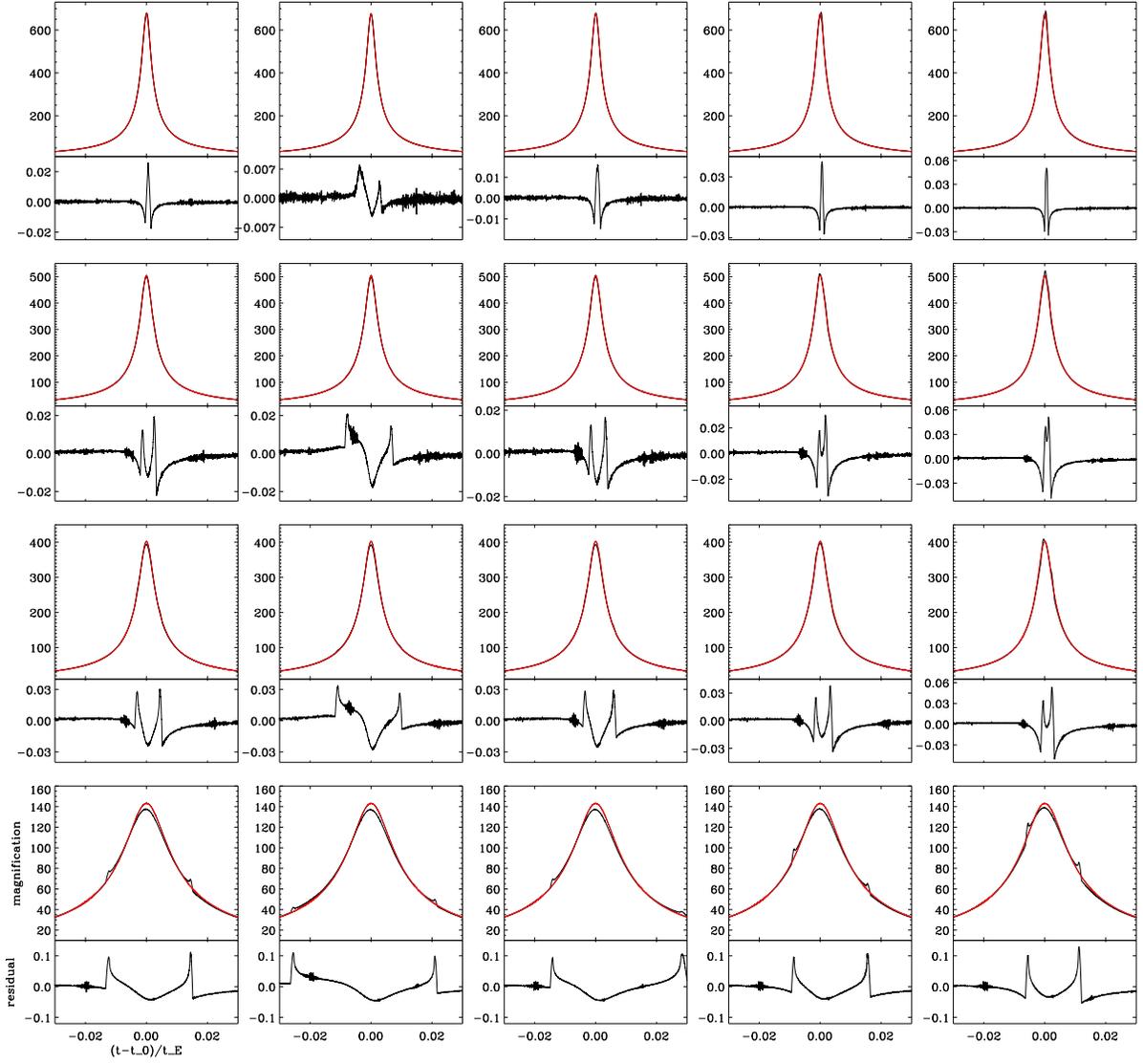}
\caption{\label{fig:two}
Light curves resulting from the source trajectories presented in Fig. 1.
In the upper part of each panel, black and red curves represent the light curves of the planetary and single lensing events, respectively.
The lower part represents the residuals from the single lensing magnification.
}\end{figure}

\begin{figure}[t]
\epsscale{1.0}
\plotone{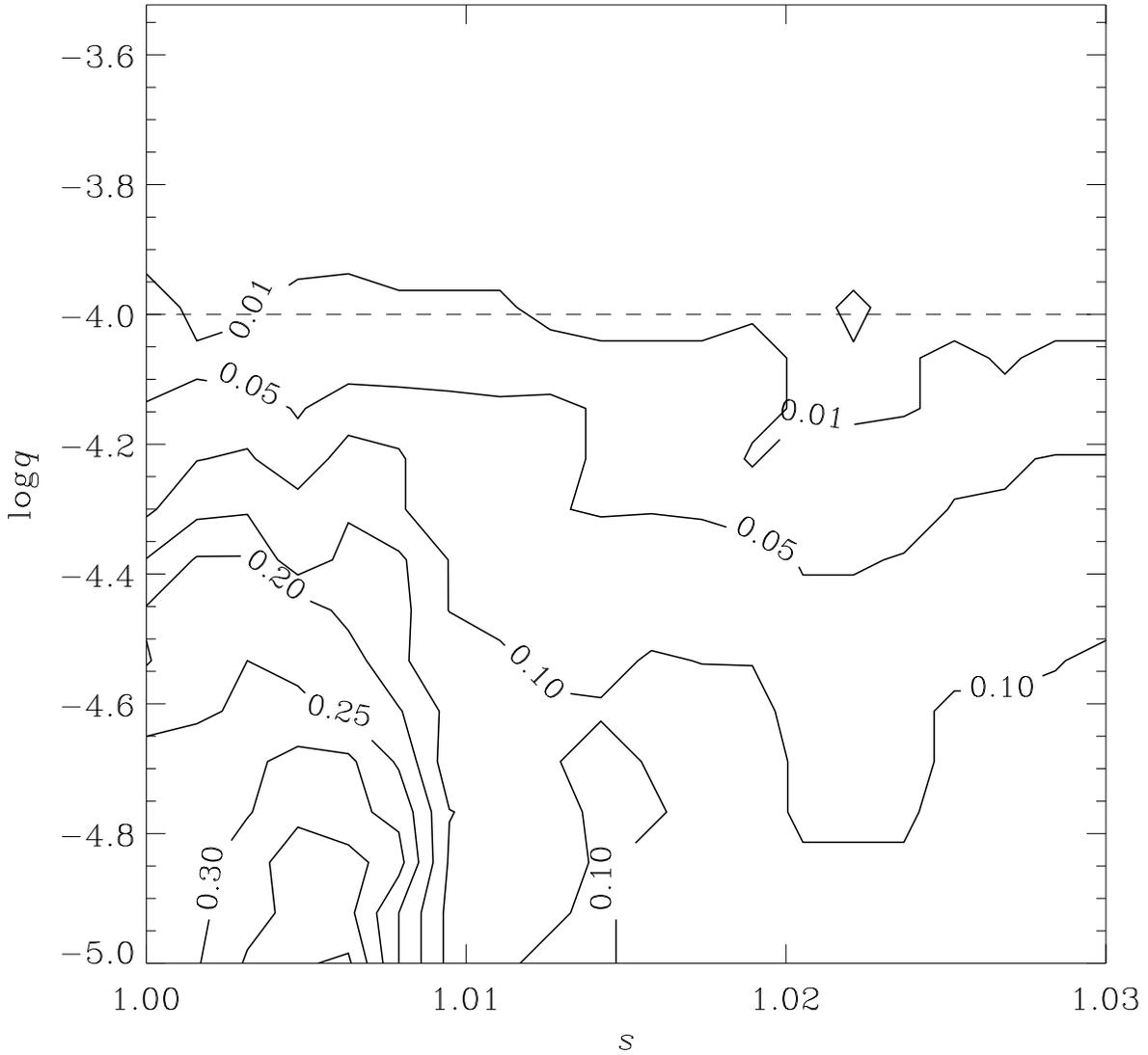}
\caption{\label{fig:three}
Probability of the occurrence of events with weak perturbation $\epsilon \leqslant 3\%$ in the resonant caustic high-magnification events as a function of the star-planet separation and planet/star mass ratio.
The dashed line indicates the mass ratio of $q=10^{-4}$.
}\end{figure}

\begin{deluxetable}{cc}
\tablecaption{Probability of the occurrence of resonant caustic events with weak perturbation.\label{tbl-one}}
\tablewidth{0pt}
\tablehead{
Planet mass & Probability ($\%$) }
\startdata
3.0$M_{E}$ & 21 \\
10.0$M_{E}$ & 8 \\
20.0$M_{E}$ & 1 \\
\enddata
\tablecomments{
The planets are located around $2.5 \rm{AU}$ from the host star by assuming a high-magnification planetary lensing event with $D_{\rm S} = 8 \ {\rm kpc}$, $D_{\rm L} = 2 \ {\rm kpc}$, and $m = 0.5\ M_\odot$.
}
\end{deluxetable}


\end{document}